\documentclass{ws-procs9x6}
\usepackage{color}

\begin{document}

\title{NEW SUB-MILLIMETER HETERODYNE OBSERVATIONS OF CO AND HCN IN TITAN'S ATMOSPHERE WITH THE
APEX SWEDISH HETERODYNE FACILITY
INSTRUMENT$^{\star}$\footnotetext{$^{\star}$Based on observations
collected at the European Southern Observatory, Chile, corresponding
to the observing proposal number 081.F-9812(A).}}

\author{M. RENGEL$^*$, H. SAGAWA$^\dagger$\footnotetext{$^\dagger$Currently at the National
    Institute of Information and Communications Technology, Japan}, and P. HARTOGH}

\address{Max-Planck-Institut f\"ur Sonnensystemforschung,\\
Max-Planck-Stra${\ss}$e 2, D-37191 Katlenburg-Lindau, Germany\\
$^*$E-mail:rengel@mps.mpg.de\\
www.mps.mpg.de/homes/rengel/}
\begin{abstract}
The origin of the atmosphere of the largest moon of Saturn, Titan,
is poorly understood and its chemistry is rather complicated.
Ground-based millimeter/sub-millimeter heterodyne
spectroscopy resolves line shapes sufficiently to determine
information in Titan's atmospheric composition (on vertical
profiles and isotopic ratios).
We test the capabilities of
the Swedish Heterodyne Facility Instrument (SHFI),
Receiver APEX-1, together with the Atacama Pathfinder EXperiment APEX 12-m
telescope for
Titan's atmospheric observations. In particular we
present sub-millimeter
observations of the CO(2-1) and HCN(3-2) lines of the Titan
stratosphere with APEX,
and with SHFI taken
during the Science Verification (SV)
instrument phase on March and June 2008.
With the help of
appropriate radiative transfer calculations we investigate the possibility to constrain the
chemical concentrations and optimize the
performance of the APEX-1
instrument for inferring vertical profiles of molecular components
of the atmosphere of Titan.
\\
This study attempts to
contribute to constrain radiative transfer and retrieval algorithms
for planetary atmospheres, and to give hints to the current and future ground and space-based data acquisition leading to a more thorough understanding of
the chemical composition of Titan.

\end{abstract}


\bodymatter

\section{Introduction}\label{intro}

Titan's atmosphere exhibits a complex photochemistry. The origin of
carbon monoxide (CO) is not well understood (whether photochemical
or primordial). Hydrogen Cyanide (HCN), the most abundant nitrile in
Titan, is a key intermediate in production of more complex
hydrocarbons and organic molecules. HCN in the Titan atmosphere has
been discovered by the infrared observations of
Voyager\,1\textcolor{black}{\cite{h81}}, and CO has been detected by
the ground-based near-infrared
observations\textcolor{black}{\cite{l83}}. Following these
detections, the molecular concentrations of CO and HCN in Titan's
atmosphere have been determined from infrared, millimeter, and
sub-millimeter observations as well as from modeling (see Tables\,2
and 3 in section\,4 for details). CO seems to be uniformly mixed in
Titan's atmosphere up to high altitudes. HCN abundances, however,
display a steeper profile with ambiguous enrichment values
(Table\,3). Already several observational data confirm that Titan's
atmospheric composition is indeed
seasonally\textcolor{black}{\cite{roe04,teanby08}}, and spatially
dependent\textcolor{black}{\cite{teanby09}}. Full behavior of the
seasonal characteristics of the spatial distribution is not
constrained yet due to the limitation in the temporal coverage of
the previous observations, and to the use of different instruments.
New sub-millimeter observations are required not only to provide new
abundance constraints and shed more light on the rate of these
seasonal variations, for example, but for support, complement and
cross-calibration the ESA's Herschel Space Observatory mission. One
of the goals of the Key Program of Herschel, entitled \textit{Water
and Related Chemistry in the Solar
System}\textcolor{black}{\cite{ha09}}, is to understand water
inventory in the Titan's atmosphere as well as distributions of
other hydrocarbons and nitriles. During the guaranteed time, line
surveys on Titan at the frequency range of 500 and 5000\,GHz are
going to be carried out using two low-to-medium resolution
spectrometers (Photodetector Array Camera and Spectrometer (PACS)
and Spectral and Photometric Imaging Receiver (SPIRE)). Our model
calculations of the synthetic spectra of Titan expected to be
observed with Herschel show that several CO and HCN lines will be
detected with PACS and SPIRE\textcolor{black}{\cite{re09}}. Because
of their low-to-medium spectral resolutions, SPIRE and PACS are not
capable of measuring the shapes of CO and HCN lines, which prevents
us from determining precisely the vertical profile of CO and HCN
mixing ratios. Therefore, high spectral resolution  observations of
CO and HCN with ground-based sub-millimeter telescopes would
significantly improve the accuracy of retrieving the CO and HCN
profiles.

The Atacama Pathfinder EXperiment (APEX) 12-m telescope is
operational since 2005\textcolor{black}{\cite{g06}}, and has been already used to the
planetary science (monitoring mesospheric winds of Venus\textcolor{black}{\cite{l08}}),
among numerous galactic and extra-galactic objects.
In spring 2008, the new APEX Swedish Heterodyne Facility Instrument
(SHFI) has been commissioned on the APEX\,12-m
telescope\textcolor{black}{\cite{vas08,nys09}}, which consists of four single-pixel
receivers (APEX-1, APEX-2, APEX-3, and APEX-T2) mounted in a single
cryostat located in the Nasmyth\,A cabin of APEX. In this report we
present a summary of our observations of CO and HCN
performed with APEX-1 during the Science Verification (SV) phase with the
aims to demonstrate the capabilities of APEX and SHFI for Titan's
atmospheric observations, and to investigate the possible retrieval
of CO and HCN abundances. We also examine the possible detection of
the millimeter/sub-millimeter rotational transition of
C$_{2}$H$_{2}$ (already observed in the mid-infrared spectra
by the Composite Infrared Spectrometer, CIRS, on
Cassini\textcolor{black}{\cite{vina07}}), and the search for one new compound,
HC$_{5}$N(83-82) which the possible presence has been suggested by
experimental results simulating Titan's atmosphere
\textcolor{black}{\cite{van95,coll95,co06}}.

The observations are described in detail in section\,2 while the
radiative transfer modeling and the data analysis are discussed in
sections\,3 and 4, respectively. A brief discussion is given in
section\,5 and a summary in section\,6.

\section{Observations}

Titan's spectra were acquired with the APEX 12-m telescope located
on Chajnantor (Chile) during 21 March, and 17--18, 23, and 27 June
2008 during SV. We used the APEX-1 receiver, operating at 211-270
GHz, which employs a superconductor-insulator-superconductor (SIS)
mixer and behaves as a single sideband receiver (SSB). For the
backend, we used the Fast Fourier Transform Spectrometer (FFTS) with
a channel separation of 122\,kHz and a bandwidth of 1\,GHz. Table\,1
summarizes the measured transitions as well as the observing days
and the precipitable water vapour (PWV). Titan was observed near the
western or eastern elongations at separation angles from Saturn
greater than 120\,$^{\prime\prime}$. Pointing and focusing of the
telescope were regularly checked scanning across Saturn in azimuth
and in elevation (APEX has a pointing accuracy of
2\,$^{\prime\prime}$ r.m.s. over sky). The apparent diameter of
Titan  was around 0.8\,$^{\prime\prime}$.

\begin{table}
\tbl{Species observed on Titan} {\begin{tabular}{@{}lccc@{}}\toprule
Species      & Frequency (GHz) & Date 2008 & PWV (mm)\\
\colrule
HCN(3-2)   &  265.886 & 21 March & \\
HC$_{5}$N(83-82) & 220.937  & 17-18 June & 3.5-4.0\\
CO(2-1)     &  230.538 & 23 June    & 0.5 \\
HCN(3-2)   & 265.886  & 23 June    & 0.5 \\
C$_{2}$H$_{2}$ J=47 $\nu$5=1 $\rightarrow$ J=46 $\nu$4=1  &
254.521 & 27 June    & 0.2 \\
\botrule
\end{tabular}
} \label{op}
\end{table}

\normalsize


Because a bug in the low-level Python ephemerides software not
previously detected, Titan was not tracked properly at the beginning
of our observations on 21 March 2008. The bug was successfully
removed by the APEX staff. These observations in particular have
helped to improve the control software. Accounting for possible
frequency errors in the line center (of a few MHz around the
respective line centers), and considering the short integration
time, we could not detect any evident signature of HC$_{5}$N(83-82)
and C$_{2}$H$_{2}$ in the recorded spectra averages of integration
times of 36 and 91 min. (r.m.s. of 0.035 and 0.03\,K), respectively.
Lower r.m.s. noise levels of around 0.01\,K (which could be achieved
with integration times of around 8 and 13\,h, respectively) would
have facilitated the possible detections of these species. Therefore
we concentrate here on the CO(2-1) and HCN(3-2) observations.
Because of the high oversubscription of observing proposals and the
uncertainties about the tracking at the beginning of the campaign,
the  initially proposed integration time of around 2 hours for
CO(2-1) and 2.65 hours for HCN(3-2), which corresponds to a r.m.s
noise level of 0.017\,K and 0.025\,K respectively, was not
available. Instead, CO(2-1) and HCN(3-2) were integrated for around
13 and 19 min, respectively. The initial reduction of each spectrum
was performed using the CLASS software package of the Grenoble
Astrophysics Group\footnote{URL:http://www.iram.fr/IRAMFR/GILDAS}.

Fig.\,1 presents the recorded Titan's spectra of the CO(2-1) and
HCN(3-2) lines. The resolution has been downgraded to about 4\,MHz
in order to increase the signal-to-noise ratio.

\begin{figure}
\begin{center}
\psfig{file=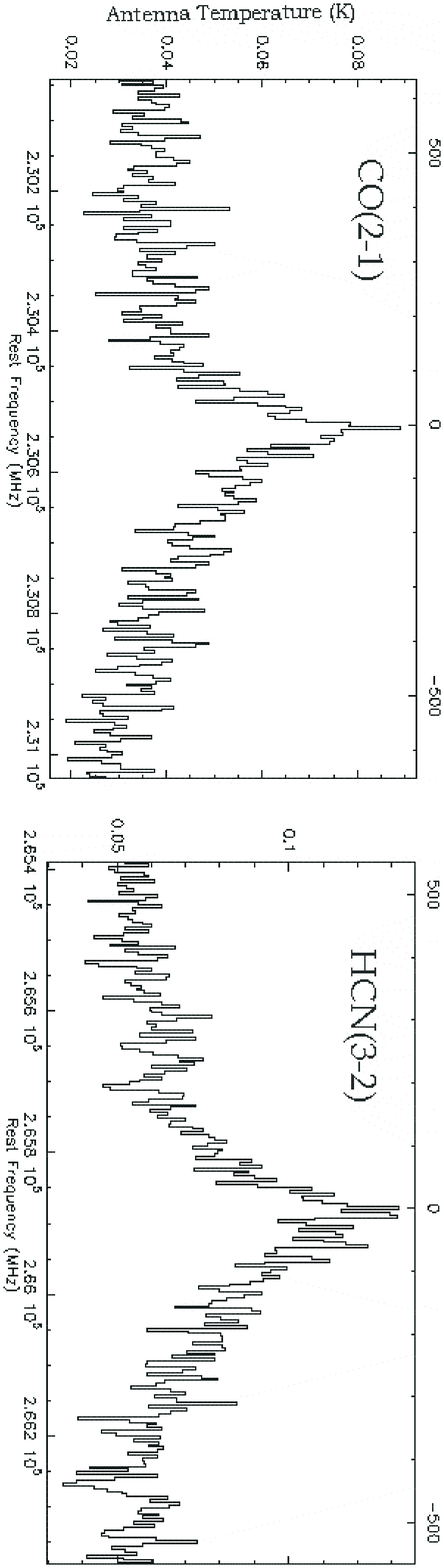,width=3
cm, angle=90}
\end{center}
\caption{Titan's spectra observed with the APEX 12-m Telescope on 23
June 2008. Left: CO(2-1) line rebinned over 32 resolution elements
(4 MHz). Right: HCN(3-2) line  rebinned over 32 resolution elements
(4 MHz).} \label{tp}
\end{figure}

The conversion from the observed antenna temperature $T_{a}^{*}$ to
the total flux density from the Titan's  atmosphere
$S_{\nu,tot}$\,[Jy] is performed by $S_{\nu,tot}$ = 24.4$\times$
$T_{a}^{*}$ $\cdot$ $\eta_f$/$\eta_a$, where $\eta_f$ and $\eta_a$
are the forward and aperture efficiencies, respectively. In this
study, we adopt the values considering the efficiencies listed on
the APEX web
site\footnote{URL:http://www.apex-telescope.org/telescope/efficiency/}
and of the instrument paper\textcolor{black}{\cite{g06}}: $\eta_a$ are 39.9 and 40.2 for 230 and 265
GHz, respectively, and $\eta_f$ is 0.95 for both frequencies. These
values are expected to have a 10\% uncertainty.

By fitting the synthetic spectra to the observed one, we intend to
infer the abundance of CO and HCN in Titan's stratosphere. A precise
comparison of the observations with synthetic spectra must cope with
two limitations: (1) the line-to-continuum ratios of these CO and
HCN spectra were not able to be confirmed from the observed data.
1\,GHz bandwidth is too narrow for constraining the continuum level.
We have examined the possibility of using the emission at the
frequency of $\pm$500\,MHz offset from the line center as the
pseudo-continuum. We found, however, that the intensities of the CO
(2-1) line at the  $\pm$500\,MHz significantly depend on the CO
abundance at a certain range of altitude, which also affects the
intensity at the line center. Therefore, we conclude that using such
fluxes as the pseudo-continuum is not a suitable method for
determining the CO abundance in this study. This is also the case
for the HCN (3-2) line.
(2) Baseline
ripples and uncorrected larger scale baseline characteristics may
affect the shape of the observed spectra. One can obtain
a spectral line shape symmetric with respect to the line center
by folding the spectrum. However, such a spectra folding may introduce an
artificial error in the line shape. Therefore we prefer to keep the
spectra with their original baseline distortions.

\section{Radiative transfer modeling}

The synthetic spectra are generated by a multi-layered line-by-line
radiative transfer model considering the spherical geometry of Titan's atmosphere. It
consists of 120 layers that span the 0--600 km interval with a
resolution of 5\,km. In general, the atmospheric model for Titan is
explained fully in Rengel et al. (2009)\textcolor{black}{\cite{re09}}. Briefly, Titan's
thermal and pressure profiles based on the in-situ measurement by the Huygens probe\cite{fu05}
are adopted for the
opacity calculation of the CO and HCN gases at each altitude level.
The mixing ratio profiles of CO and HCN are set as the parameters to be retrieved in this study. The detail of modeling vertical profiles of each molecule is described later.
For CO, the assumed N$_2$-broadening line-width here is  that of Semmoud-Monnanteuil \& Colmont\,(1987)\textcolor{black}{\cite{sc87}}, and for HCN, the value corresponding to the same transition at the infrared vibrational-rotational band\cite{smith84} as there is no exact measurement for this pure rotational line.
The collision
induced absorption coefficients of N$_2$, CH$_4$, and H$_2$
mixtures, which dominate the millimeter and sub-millimeter continuum
opacity of Titan's atmosphere, are included by using the formulation
of Courtin (1988)\cite{c88} and Borysow and Tang (1993)\textcolor{black}{\cite{bo93}}.

\section{Data analysis}

\subsection{The CO abundance}

Recent observations suggest that CO in Titan is uniformly mixed throughout Titan's atmosphere\textcolor{black}{\cite{gu04,ba06,kok07}}. Such a profile can be supported if its long chemical lifetime in the oxygen-poor Titan atmosphere is considered ($\sim$10$^{-9}$ years\textcolor{black}{\cite{yu84}}). In this study, we retrieve CO abundance by assuming a vertically constant profile. It should be noted that some of the previous observations have shown the possibility of existence of the discontinuity in the CO mixing ratios between the troposphere and the stratosphere (e.g., $\sim$30\,ppm in the troposphere\cite{le03} and $\sim$60\,ppm in the stratosphere\textcolor{black}{\cite{lv05}}). However, such discontinuity cannot be constrained by our observations, as the CO(2-1) measurement reported here does not have any sensitivity to the altitude below 60\,km (Fig.\,2 left panel).

A direct comparison of synthetic and observed spectra is made and
the best fitting model with the least $\chi^{2}$ is determined. In Fig.\,2 right panel, the best-fit CO abundance shows a
solution of 30$^{+15}_{-8}$\,ppm.
Although this solution agrees with the previous stratospheric CO measurements
of Hidayat et al. (1998)\textcolor{black}{\cite{hi98}}, Lellouch et. al
(2003)\textcolor{black}{\cite{le03}}, and Baines et
al. (2006)\textcolor{black}{\cite{ba06}} (Table\,2), we have to allow a larger
error on our result by considering
the uncertainty in the flux scaling and the baseline problems.
To examine the effect of the uncleaned baseline of the observed spectrum, we performed the fitting analysis by constraining the data only on each side of the wings. Using the red-shift wing ($\nu$ $<$ $\nu$$_{0}$) yielded the best fitting CO mixing ratio of 26$^{+13}_{-8}$\,ppm, while using the blue-shift wing ($\nu$ $>$ $\nu$$_{0}$) resulted 36$^{+15}_{-11}$\,ppm.
Taking into account the 10\% uncertainty in the flux calibration, the acceptable range of CO mixing ratio
becomes as 16--51\,ppm. This wide range covers most of the past works. One way
to avoid such a large ambiguity in the data analysis
(i.e. to avoid the effect of the uncertainty in the flux calibration) is to
calibrate the observed spectrum by scaling its continuum level to that of the forward model calculation. Our forward model calculations suggest that the intensity at the frequency offset larger than $\sim$1.5\,GHz can be used as the continuum level.

\begin{figure}
\begin{center}
\psfig{file=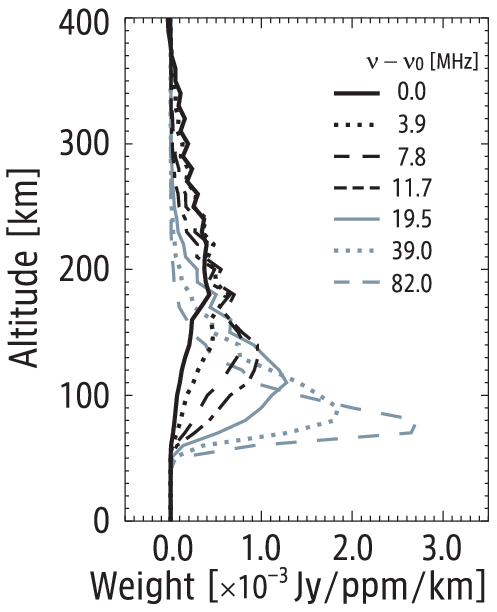,width=4 cm, angle=0}
\psfig{file=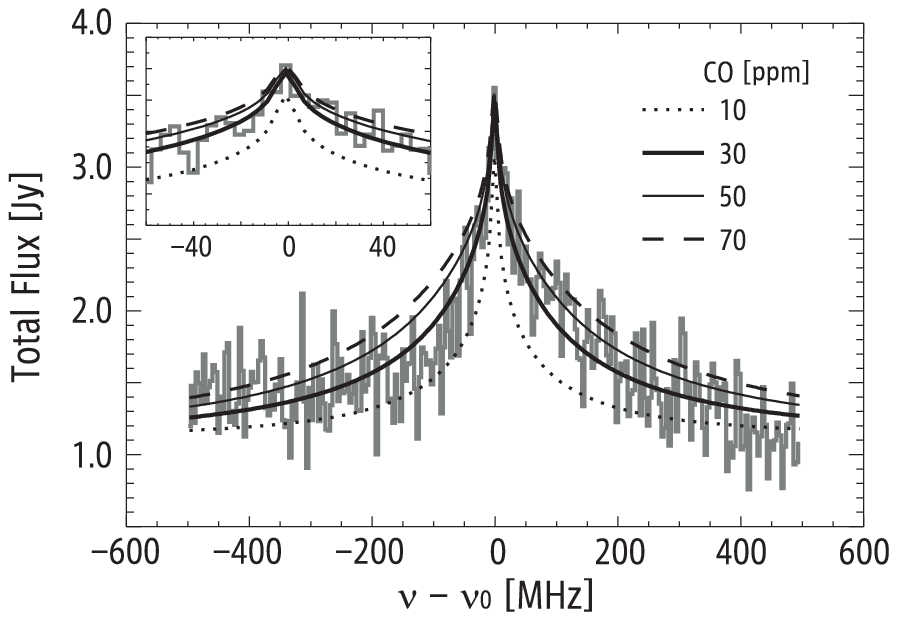,width=7 cm, angle=0}
\end{center}
\caption{Left: Weighting functions with respect to CO mixing ratio for selected frequencies near the CO(2-1) line center. Note that the finite spectral resolution (4\,MHz) of the measurement is taken into account for the calculation of these weighting functions. Right: Observed and synthetic CO(2-1) spectra at 230.54\,GHz (grey spectrum and line ones, respectively). The synthetic spectra are calculated assuming vertically uniform CO mixing ratios of
10, 30, 50, and
70\,ppm (from bottom to top) respectively. The small upper panel shows a zoom of the spectrum for a frequency range of $\pm$60\,MHz
. } \label{tp}
\end{figure}

\begin{table}
\tbl{Observations of CO in Titan atmosphere}
{\begin{tabular}{@{}ccccc@{}}\toprule
Altitude   &Mixing ratio [ppm]&   Wavelength$^*$ [$\mu$m]& Facility/Instrument & Reference\\
\colrule
Troposphere & 48$^{+100}_{-32}$ & 1.57  & KPNO-4m/Fourier Transform interferometer & \cite{l83}\\
Stratosphere & 30--180 & 35314.89 & VLT-70m/27 antennas & \cite{m84}\\
Stratosphere & 60$\pm$40 & 2602.59 & OVRO/2 elements interferometers & \cite{pa84}\\
Stratosphere & 2$^{+2}_{-1}$ & 2602.59 & IRAM-30m/3-mm\,SIS receiver & \cite{mar88}\\
Stratosphere & 50$\pm$10& 2602.59 & OVRO/6 10.4\,m diam. antennas & \cite{gm95}\\
Troposphere & 10$^{+10}_{-5}$ & 4.8 & UK IR Telescope/CGS4 spectrograph & \cite{noll96}\\
Stratosphere &27$\pm$ 5 & 2602.59, 1304.34, 900.9 & IRAM-30m/SIS heterodyne receivers& \cite{hi98}\\
Stratosphere &52$\pm$6 & 1304.34 & OVRO/6 antennas & \cite{gu00}\\
Troposphere & 32$\pm$10 & 4.75--4.85& VLT\,8-m/ISAAC & \cite{le03}\\
Stratosphere & 51$\pm$4 & 869.56  & SMA/5 and 6 antennas & \cite{gu04}\\
153--350\,km & 60 & 4.5-4.85 & VLT\,8-m/ISAAC  & \cite{lv05}\\
Tropo-Stratosphere& 45$\pm$15 & 4.64 & Cassini/CIRS & \cite{fla}\\
Stratosphere & 32$\pm$15 & 4.64 &  Cassini/VIMS & \cite{ba06}\\
Stratosphere & 47$\pm$8 & 333.3-166.66 & Cassini/CIRS & \cite{dekok07}\\
Stratosphere & 30$^{+15}_{-8}$  & 1301.29 & APEX/APEX-1 receiver & This article\\
\botrule
\end{tabular}} \label{co}
$^*$In spite that each wavelength region is defined by standard specifical units
, we use here $\mu$m to facilitate the comparisons.
\end{table}

\normalsize

\subsection{The HCN vertical profile}
We retrieve the mixing ratio profile of HCN by performing forward- and inversion calculations iteratively. For solving the inversion problem, we employ the optimal estimation method (OEM) (e.g.,
see details in Rodgers\,(1976)\textcolor{black}{\cite{ro76}}). The
  Levenberg-Marquardt scheme is used in the iteration.
We set the logarithm of HCN mixing ratio as the retrieving parameter.
The \textit{a priori} profile of HCN mixing ratio, which is required in the OEM approach to regularize the ill-posed inversion problem, is based on the previous ground-based work by Marten et al. (2002)\textcolor{black}{\cite{ma02}}.  We assumed the \textit{a priori} error of 2.3 in the logarithm scale.
During the HCN retrieval, the temperature profile is fixed to our nominal temperature profile (the one based on the Huygens measurements).

Fig.\,3 shows the best-fit synthetic spectrum (solid line) and the corresponding HCN profile. The averaging kernels of the retrieved HCN mixing ratios indicate that this retrieval has a significant sensitivity at the altitude around 90\,km though we can still retrieve information at the higher altitudes up to 240\,km.
Our retrieved HCN mixing ratios are larger than those of Marten et al (2002) by a factor of $\sim$30--80\% at the altitudes of 100--200\,km. The synthetic spectrum calculated with the \textit{a priori} HCN profile, i.e., the result of Marten et al. (2002), is also shown for comparison (Fig.\,3). A large discrepancy between the observed and the synthetic spectra, and the \textit{a priori} profile is seen mostly at the wings of the spectrum. This results in the retrieved larger HCN abundances at the altitudes of 90--150\,km. Note that our result showing lower HCN abundance values at around 300\,km is out of the altitude range where our observations are sensitive.

In the above mentioned retrieval, we only took the system
temperature of the APEX-1 instrument into account for the
statistical error of the measurement. In fact, as we discussed
previous in section\,4.1, the observed spectrum contains the 10\%
uncertainty in the flux calibration. If a 10\% larger scaling factor
is applied in the flux calibration, we cannot derive any realistic
HCN profiles which can reproduce the observed spectrum. In case of
using a 10\% smaller scaling factor, the HCN abundances for the
best-fit spectrum are larger than those of Marten et al. (2002) at
the altitude of 100--150\,km (which is attributed by the wings of
the spectrum) and then becomes as small as $\sim$30\% of Marten et
al. (2002) at higher altitudes of 150--200\,km (Fig.\,4).

The two retrievals (Figs.\,3 and 4) demonstrate the significant impact of the flux calibration uncertainty on the retrieved HCN profile. Since we cannot constrain the line-to-continuum ratio of the observed spectrum, we are not able to conclude the HCN abundance at altitude of 150--200\,km which deviates within a wide range from 30 to 150\% of the \textit{a priori} guess.
For the lower altitudes such as 100--150\,km, the wing of our observed spectrum requires a higher opacity of HCN than that of Marten et al. (2002) even if we consider the uncertainty in the flux calibration.


\begin{figure}
\begin{center}
\psfig{file=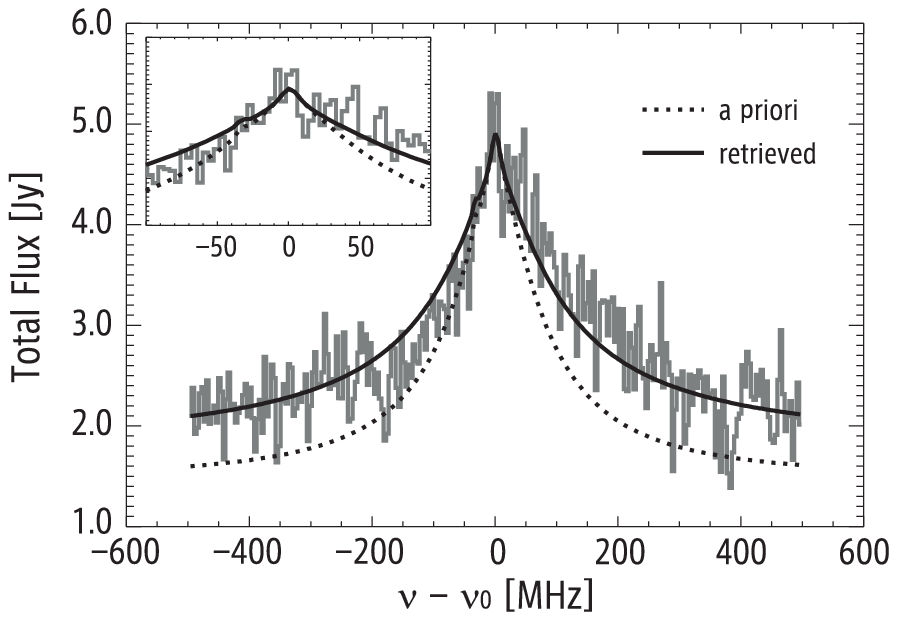,width=5.5 cm, angle=0}
\psfig{file=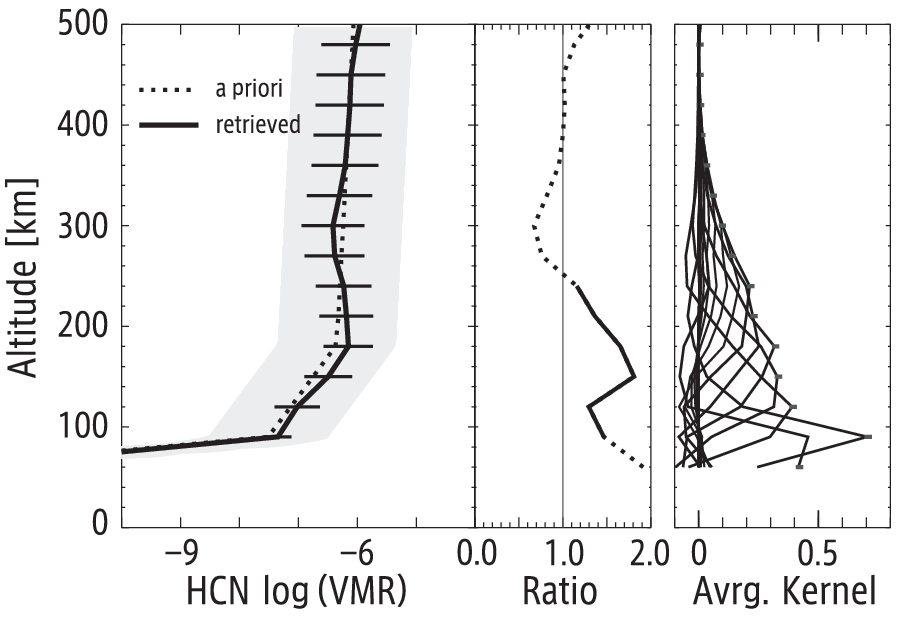,width=5.5 cm, angle=0}
\end{center}
\caption{Left: Observed and synthetic HCN(3-2) spectra at
265.89\,GHz: Solid line
 shows the best-fit spectrum, and dash line the spectrum calculated with the
\textit{a priori} HCN profile.
 Grey line is the observed data. The upper panel figure represents the zoom to the frequency range of +/-100\,MHz. Right:
The retrieved HCN vertical profile (solid line) under the nominal temperature profile without considering the error in the flux calibration. The error bar indicates the 1-$\sigma$ limit of the retrieval. The dotted profile represents the \textit{a priori} values where the gray shaded region corresponds to their errors. The ratio of the retrieved profile with respect to the \textit{a priori} profile is also shown. On the extrem right panel, the averaging kernels of the retrieved HCN mixing ratios are shown.
} \label{tp}
\end{figure}

\normalsize

\begin{figure}
\begin{center}
\psfig{file=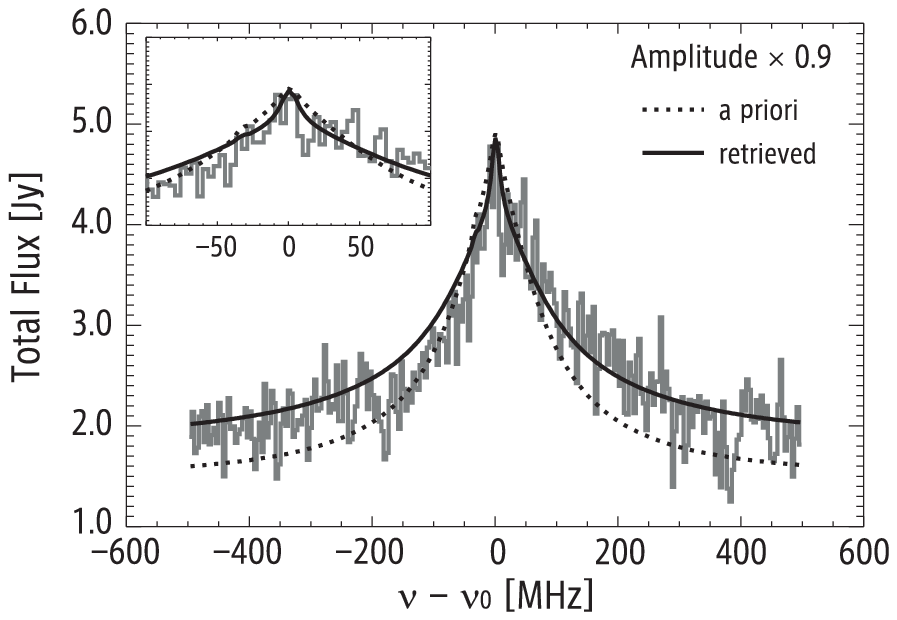,width=5.5 cm, angle=0}
\psfig{file=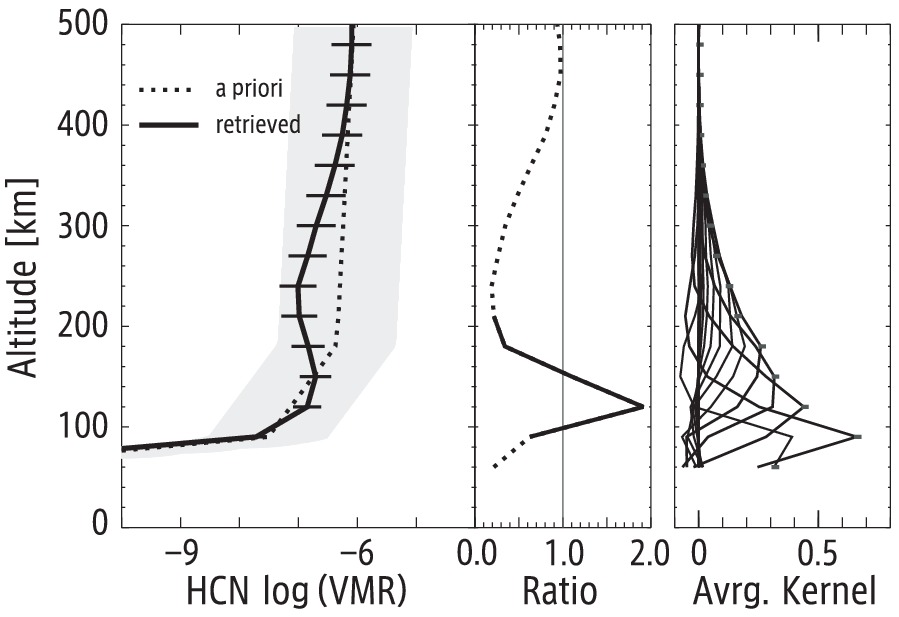,width=5.5 cm, angle=0}
\end{center}
\caption{See caption Fig. 3, but here the measurement data is scaled by a factor of 0.9.
} \label{tp}
\end{figure}

\begin{table}
\tbl{HCN mixing ratios in Titan's atmosphere} {\begin{tabular}{@{}ccccc@{}}
\toprule
Altitude  &Mixing ratio [ppm]&   Wavelength$^*$ [$\mu$m]& Facility/Instrument & Reference\\
\colrule Stratosphere & 3.0 $\times$ 10$^{-7}$ & 3386.0 & IRAM-30m/SIS receiver
& \cite{pa87}\\
Stratosphere & 1.6 $\times$ 10$^{-7}$ &  14.025 & Voyager/IRIS & \cite{c89}\\
Stratosphere & (0.75--52)$\times$ 10$^{-7 }$ & 3386.0 & IRAM 30-m/SIS receiver &
 \cite{t90}\\
Stratosphere & 4.7$\times$ 10$^{-8}$--1.5 $\times$ 10$^{-6}$ & 14.025 &Voyager/
IRIS & \cite{cb95}\\
400\,km& $\sim$ 2 $\times$ 10$^{-5}$  &  & Model prediction&\cite{la96}\\
Stratosphere  &(0.5--4) $\times$ 10$^{-7}$ & 14.025 & IRAM 30-m/SIS receiver
&\cite{h97}\\
400\,km & 1 $\times$ 10$^{-6}$ & & Model prediction & \cite{l01}\\
700\,km & 1 $\times$ 10$^{-5}$ & & Model prediction & \cite{l01}\\
400\,km & 1 $\times$ 10$^{-6}$ &  3386.0 & IRAM-30m/SIS receiver & \cite{mar02}\\
400\,km & 1 $\times$ 10$^{-5}$ & & Model prediction & \cite{g03}\\
700\,km &1 $\times$ 10$^{-4}$ & & Model prediction & \cite{g03}\\
Stratosphere  & 3.0 $\times$ 10$^{-7}$ & 14.025 &   ISO/SWS & \cite{c03}\\
$\sim$600\,km & 7 $\times$ 10$^{-3}$ & 3 & Keck II/NIRSPEC  & \cite{g03}\\
83\,km &  3$\times$ 10$^{-5}$ & 1692.43 & SMA/5 and 6 antennas & \cite{gu04}\\
300\,km &  3 $\times$ 10$^{-5}$ & 1692.43 & SMA/5 and 6 antennas & \cite{gu04}\\
500\,km & 0.4 $\times$ 10$^{-6}$ & 14.04 & Cassini/CIRS & \cite{ta07}\\
\botrule
\end{tabular}} \label{h}
$^*$In spite that each wavelength region is defined by standard specifical units
, we use here $\mu$m to facilitate the comparisons.
\end{table}

\begin{figure}[h]
\begin{center}
\psfig{file=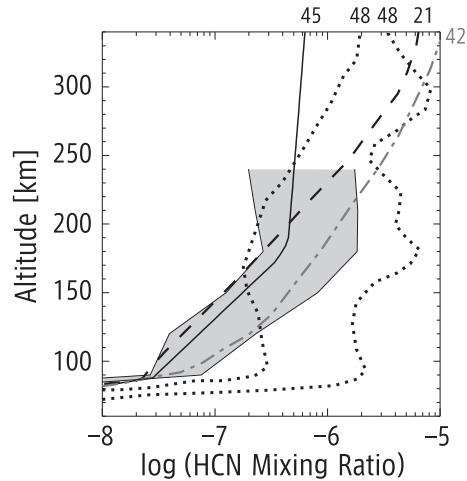}
\end{center}
\caption{Comparison of recent observational and modeling works of
HCN abundances. The small numbers correspond to the indices of the
reference on Table\,3. The shaded region represents the results here
obtained, considering the 1-$\sigma$ limit of the retrieval error.}
\label{HCNprof}
\end{figure}

\section{Discussion}

How does one explain the behavior of the retrieved HCN vertical profile
obtained here? the relative poor signal-to-noise of the observations
obtained here
results in a poor fit to the HCN line wings. Furthermore,
comparing the results obtained with previous results
derived from disk-averaged information, interferometer measurements,
and space-based observations summarized in Table\,3 must be
considered cautiously. In other words, direct comparison of the
results from different instruments requires us to assume that all
instrument-related offsets have been accounted for.

Certainly, Titan's atmospheric temperature and composition are not
in a steady state. Does the HNC abundance retrieved here perhaps
indicate that the composition of Titan's atmosphere has changed
during the last years? Fig.\,7 in Teanby et
al.\,2009\textcolor{black}{\cite{teanby09}} shows that the HCN
abundance measured by Cassini (at southern latitudes) has not
significantly changed between 2006 and 2008. Then it seems to be
that possible detections of time variations (if any) between
different observations could be maybe due to observing more of the
northern hemisphere than of the southern one as Titan's season
changes. As we derived disk-averaged information we need to keep in
mind, however, that our vertical distributions are mostly
representative of the equatorial region of Titan since the measured
flux density spectra are more heavily weighted toward equatorial
latitudes. In case that vertical distributions would be
representative of the northern latitudes, the profile analysis would
become rather complicated. There is evidence of a vortex, which acts
to separate enriched from unenriched air, at $25-55\,^{\circ}$\,N
encircling the North Pole
\textcolor{black}{\cite{teanby09,fla,ach08}}.

Further higher signal-to-noise HCN observations at different times
are necessary in order to open the possibility to detect seasonal and
even spatial variations in HCN, and confirm or not an enrichment
layer in Titan's stratosphere.

\section{Summary}
\begin{itemlist}
\item We report the first observations obtained with the SFHI APEX-1 instrument
on APEX on a planetary/satellite atmosphere taken during SV. It
consists of the spectra of CO(2-1) and HCN(3-2) in Titan's
atmosphere.

\item The observations reported here improved the control software of the APEX
telescope,  so that now it is possible to track planets with the APEX-1
receiver.

\item We investigate the CO and HCN composition of Titan's stratosphere.
Our CO mixing ratio estimation is consistent with other authors. We
retrieved the HCN vertical profile, which is inconsistent with
previous analyses. This  requires further investigation by future
HCN observations with higher signal-to-noise ratios. The search of nitriles
and CO appears favorable in the sub-millimeter range explored
with the APEX telescope and with the APEX-1 receiver.
\end{itemlist}

This work represents a first step in exploring the capabilities of
APEX and SHFI for Titan's atmospheric observations. New simultaneous
observations of HCN and its isotopes at higher frequencies with
APEX-2 and APEX-T2, for example, would improve the sensitivity to
the abundance retrievals, and potentially could support, complement,
and cross-calibrate the ESA's Herschel Space Observatory mission.

\section*{Acknowledgments}
We are grateful to the APEX staff, in particular to Carlos De Breuck
(ESO), for its assistance. We thank the anonymous reviewers for their constructive comments. This publication is based on data
acquired with the Atacama Pathfinder Experiment (APEX). APEX is a
collaboration between the Max-Planck-Institut f\"ur Radioastronomie,
the European Southern Observatory, and the Onsala Space Observatory.


\bibliographystyle{ws-procs9x6}
\bibliography{titan-rsh-reresubmission-v1}

\end{document}